# Surfactant-aided exfoliation of molydenum disulphide for ultrafast pulse generation through edge-state saturable absorption

Richard C.T. Howe[x,1], Robert I. Woodward[2], Guohua Hu[1], Zongyin Yang[1], Edmund J. R. Kelleher[2], Tawfique Hasan[*,1]

[1] Cambridge Graphene Centre, University of Cambridge, 9 JJ Thomson Avenue, Cambridge CB3 0FA, UK
[2] Femtosecond Optics Group, Department of Physics, Imperial College London, SW7 2AZ, UK

[x] Corresponding author: e-mail rcth2@cam.ac.uk, Phone: +44 1223 7 62381
[*] Corresponding author: e-mail th270@cam.ac.uk, Phone: +44 1223 7 48362

We use liquid phase exfoliation to produce dispersions of molybdenum disulphide ($MoS_2$) nanoflakes in aqueous surfactant solutions. The chemical structures of the bile salt surfactants play a crucial role in the exfoliation and stabilization of $MoS_2$. The resultant $MoS_2$ dispersions are heavily enriched in single and few (<6) layer flakes with large edge to surface area ratio. We use the dispersions to fabricate free-standing polymer composite wideband saturable absorbers to develop mode-locked and Q-switched fibre lasers, tunable from 1535-1565 and 1030-1070 nm, respectively. We attribute this sub-bandgap optical absorption and its nonlinear saturation behaviour to edge-mediated states introduced within the material band-gap of the exfoliated $MoS_2$ nanoflakes..

**1 Introduction** Transition metal dichalcogenides (TMDs) are one of the largest families of 2 dimensional (2d) materials, with around 40 different forms, including metallic, semiconducting and insulating materials [1]. TMDs have general formula $MX_2$, where M is a transition metal (e.g. Molybdenum, Tungsten etc), and X is a chalcogen i.e. a group VI element (e.g. Sulphur, Selenium etc). Their general structure is made up from quasi-2d layers bound together by van der Waals forces, with each layer containing a plane of metal atoms covalently bonded between two planes of chalcogen atoms. Semiconducting TMDs are of great interest for photonic and optoelectronic applications, with a range of bandgaps spanning the visible spectrum [1,2]. In particular, in few-layer samples (i.e. <10 layers), the bandgaps are highly dependent on the number of layers, typically shifting from indirect in bulk materials to direct in monolayers [1].

Among the TMDs, molybdenum disulphide ($MoS_2$), with a 1.29 eV (961 nm) indirect bandgap in bulk and 1.8 eV (689 nm) direct bandgap in monolayers [3], is perhaps the most intensively studied 2d material aside from graphene. Indeed, reports on the optical properties of few-layer $MoS_2$ date back as far as the 1960s [4], while monolayer $MoS_2$ was reported in 1986 [5]. Although applications of this 2d material were not explored at that time, the recent emergence of graphene has led to a renewed interest in $MoS_2$ for electronics and photonics.

One such potential application is in ultrafast optical pulse generation. Short optical pulses are in high demand in fields as diverse as spectroscopy and manufacturing [6,7]. A major way to produce them is through fast optical switches such as saturable absorbers (SAs). SAs exhibit decreasing optical absorbance under increasing illumination and can dynamically modulate the gain characteristics of a laser cavity to generate a pulsed output via techniques such as mode-locking and Q-switching [8]. The currently dominant SA technology is semiconductor saturable absorber mirrors (SESAMs), which consist of single or multiple III-V semiconductor quantum wells. However, limitations such as narrowband tunability and complex manufacturing processes have driven research towards alternatives, with 1d and 2d nanomaterials such as carbon nanotubes (CNTs) and graphene emerging as promising candidates [9–11]. More recently, semiconducting 2d nanomaterials such as MoS2 [12], MoSe2 [13], WS2 [14], and black phosphorus [15] have also attracted significant interest in ultrafast photonics, due to the optical nonlinearity and ultrafast carrier dynamics.

MoS$_2$ possesses key properties required for an effective SA. At visible wavelengths, the third-order nonlinear susceptibility of solution processed MoS$_2$ exceeds that of graphene produced via the same method [16]. This allows for large modulation depths (i.e. the difference between absorbance in the saturated and unsaturated regimes). MoS$_2$ shows ultrafast carrier dynamics [17–19], with intraband relaxation of 30 fs [17] and carrier life-times of ~100 ps [19], comparable to those of graphene [20], allowing for fast switching and recovery. Finally, the bandgap of MoS$_2$ could allow for operation at visible wavelengths, something yet to be demonstrated using graphene or CNTs [10].

Single- and few-layer MoS$_2$ can be produced either by bottom-up methods such as chemical vapour deposition [21,22], or top down techniques such as mechanical cleavage [3] and solution-based exfoliation [23]. The latter is of particular interest, as it is a scalable technique, can be carried out under ambient conditions, and produces dispersions of single- and few-layer flakes, either through chemical exfoliation techniques such as intercalation [24], or through ultrasound or shear-mixing assisted techniques such as liquid phase exfoliation [23,25].

In addition to being suitable for low temperature, upscalable processing, solution-exfoliated MoS$_2$ can be coated/printed onto substrates, filtered to form films or blended with polymers to produce composites, making it suitable for a wide range of applications. Indeed, solution-exfoliated MoS$_2$ has allowed integration into photonic devices via polymer composite [26–32] or by transfer of coated/vacuum filtered films on to optical components such as quartz substrates, mirrors and between fibre connectors [33–35]. Additionally, small flakes of MoS$_2$ such as those produced by solution exfoliation exhibit increased optical absorption, even at energies smaller than the material bandgap. It has been suggested that this sub-bandgap absorption, unexpected in perfect MoS$_2$ crystals, results from edge states, similar to an earlier observation on lithographically patterned/micro crystals of MoS$_2$ with high edge to surface area ratio [36]. Solution exfoliated MoS$_2$ is therefore an attractive material for wideband pulse generation in the infrared and potentially, in the visible region, with recent results demonstrating pulse generation from 1030 to 2100 nm, in both discrete and tunable setups [26–35, 37–44]. Similarly broad tunability has been demonstrated with other TMDs such as molybdenum diselenide (MoSe$_2$) [13], and tungsten diselenide (WS$_2$) [14], suggesting that TMDs form a diverse group of materials for ultrafast pulse generation.

Here we present the fabrication of a MoS$_2$-polymer composite SA based on ultrasonic assisted liquid phase exfoliation (UALPE). We discuss the role of surfactant chemical structure in exfoliation and stabilisation of the MoS$_2$ flakes. The MoS$_2$-SA is used to mode-lock and Q-switch fibre lasers at 1535-1565 and 1030-1070 nm, respectively, demonstrating the potential of solution processed MoS$_2$ for ultrafast photonics. We suggest that this sub-bandgap operation arises from edge-state absorption.

## 2 Exfoliation and characterisation of MoS$_2$

**2.1 Liquid phase exfoliation** The UALPE process for MoS$_2$ is similar to that for other layered materials [23] and consists of two steps. First, MoS2 bulk crystals are mixed with a suitable solvent and exfoliated via mild ultrasonication. The ultrasound causes high-frequency pressure variations and formation of microcavities in the solvent. Collapse of these cavities produces high shear-forces, exfoliating flakes from the bulk crystals by overcoming interlayer van der Waals forces. The second step is to remove unexfoliated flakes, typically via centrifugation [23].

Like other layered materials, stabilisation of exfoliated MoS2 flakes in solvents is governed by minimised enthalpy of mixing ($\Delta H$) [11,45]. It has been demonstrated that this can be achieved by matching the polar ($\delta_P$), hydrogen bonding ($d_H$) and dispersive ($d_D$) Hansen solubility parameters of the solvent to those experimentally derived for MoS$_2$ ($\delta_{P;MoS2}$ ~9 MPa$^{1/2}$, $\delta_{P;MoS2}$ ~7.5 MPa$^{1/2}$, $\delta_{P;MoS2}$ ~17.8 MPa$^{1/2}$) [46]. Using this criterion, nonaqueous solvents such as N-methyl-2-pyrrolidone (NMP) and N-cyclohexyl-2-pyrrolidone (CHP) were identified as 'good' solvents for MoS$_2$ [46]. However, many of these solvents present processing challenges due to their high boiling points (e.g. NMP ~202°C, CHP ~154°C). UALPE in pure, lower boiling point solvents such as water and alcohols is typically not possible since large differences in between these solvents and MoS$_2$ leads to high $\Delta H$, and thus, unstable dispersions. Like nanotubes and graphene, the alternative is to use dispersants such as surfactants with these solvents [10,45]. Surfactants act by adsorbing on to the MoS$_2$ surface, reducing the interfacial tension between MoS$_2$ and the solvent. For the case of ionic surfactants, they may additionally stabilise the dispersion by inducing a net surface charge around the exfoliated MoS$_2$ flakes, generating a Coulomb repulsion to prevent reaggregation [11]. Since the presence of surfactant required for exfoliation does not influence the optical properties of SAs, exfoliation into water represents an attractive route to MoS2-based SA fabrication. Indeed, this approach remains the most widely used for 1 and 2d material based SAs [11,12,47,48].

Previous studies with graphene and carbon nanotubes have shown that some of the most effective surfactants for exfoliation and dispersion stabilisation are facial amphiphiles (i.e. molecules with a quasi-flat molecular structure with hydrophobic and hydrophilic faces), including bile salts such as sodium cholate (SC), sodium deoxycholate (SDC) and sodium taurodeoxycholate (TDC) [10, 11]. SC and SDC have very similar structures, consisting of a rigid cyclopentenophenanthrene nucleus with hydrophilic hydroxyl groups (-OH) on one face (commonly termed α), hydrophobic methyl groups (-CH$_3$) on the other (β) face, and an aliphatic chain terminating in a hydrophilic group [49], as shown in Figure 1. SDC has two -OH groups on its α face, while SC has three, meaning that SDC has a higher hydrophobic index than SC [11], and would be expected to adsorb more strongly onto the hydrophobic $MoS_2$ [50] surface. Working with two such similar surfactant molecules allows us to directly study the effect of surfactant hydrophobic index on exfoliation by excluding other variables.

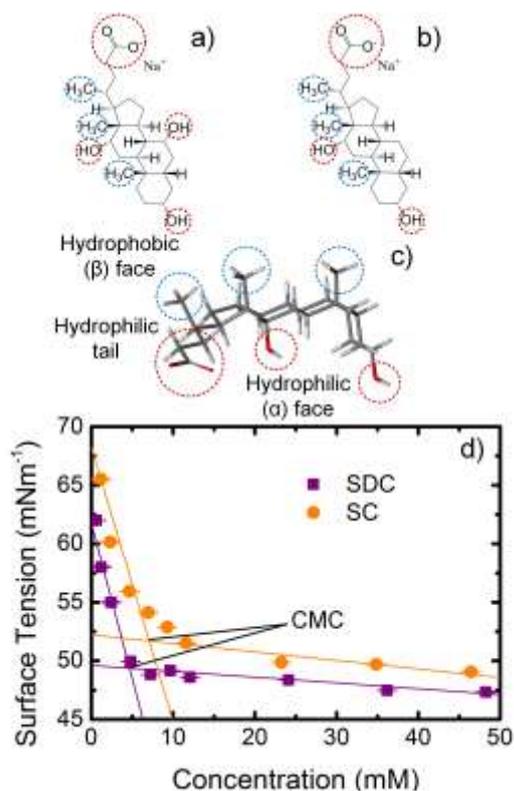

**Figure 1** Structures of bile salts SC (a) SDC (b 2d, c 3d). The hydrophilic (red) and hydrophobic (blue) groups are highlighted. Calculation of the critical micelle concentration of SC and SDC in DI water at room temperature (d).

The required concentration of surfactants for stable dispersion may be estimated by considering the surfactant critical micelle concentration (CMC). Above the temperature dependent CMC, the surfactant molecules spontaneously arrange into micelles in water [51]. Thus, if exfoliated flakes are present in the dispersion, the surfactant molecules can encapsulate and stabilize them, preventing reaggregation [10,45,52]. The CMC of a surfactant at a specific temperature may be estimated by measuring the surface tension of its aqueous solution against surfactant concentration [51]. Below the CMC, addition of surfactant causes a large change in the surface tension as the surfactant molecules assemble at the water-air interface, while above, the interface is saturated, and micelles are formed causing only a small change in surface tension [51]. We measure the surface tension for different SC and SDC concentrations in DI water via a pendant droplet method; see Figure 1(d). The measured CMC is 4.7 mM, (~2 $gL^{-1}$) for SDC and 6.9mM (~3 $gL^{-1}$) for SC. We select a surfactant concentration of 10 $gL^{-1}$, >3 times the CMC to support $MoS_2$ exfo-liation. We note that the small difference in SDC and SC molar concentration is unlikely to influence the resultant $MoS_2$ concentration as the corresponding change in surface tension above CMC is very small.

The $MoS_2$ dispersions are prepared by mixing 100 mg of MoS2 bulk crystals (~6 μm) in 10 mL aqueous solution of SC or SDC. The mixtures are sonicated for 12 hours at ~15°C in a bath sonicator. The resultant dispersions are centrifuged for 1 hour at 5 krpm (~4200 $g$). The upper 80% is decanted for characterisation and composite fabrication.

**2.2 Exfoliated $MoS_2$ Characterisation** The concentration of the dispersed MoS2 can be estimated via optical absorption spectroscopy using the Beer-Lambert law $A_\lambda=\alpha_\lambda cl$, where $c$ is the concentration ($gL^{-1}$), $l$ is the distance through the dispersion that the light passes (m), and $A_\lambda$ and $\alpha_\lambda$ are, respectively, the absorption (A.U.) and material dependent absorption coefficient ($Lg^{-1}m^{-1}$) at wavelength λ (nm). The optical extinction spectrum of the dispersed $MoS_2$ is shown in Figure 2a. We note that the extinction is distinct from the absorption, as it also includes effects such as scattering [53]. We thus dilute the dispersion to 10% v/v for measurement, in order to minimise scattering contributions to the measured absorption [53]. Note, however, that our setup can only measure scattering in thin films, and not in liquids. The spectrum shows 4 characteristic $MoS_2$ peaks, the A and B excitonic peaks at ~605 nm and ~665 nm, and broad C and D peaks resulting from the transitions between higher density of state regions of the $MoS_2$ band structure at ~440 nm and ~395 nm [54]. Using $\alpha_{605}$ =1583 $Lg^{-1}m^{-1}$, $\alpha_{665}$=1284 $Lg^{-1}m^{-1}$, and $\alpha_{800}$=324 $Lg^{-1}m^{-1}$ [26], and thus considering both peaks and featureless regions of the spectrum, we estimate the concentration of dispersed $MoS_2$ to be ~0.085 $gL^{-1}$ with SDC, and ~0.051 $gL^{-1}$ with SC. This confirms that SDC is ~68% more effective in dispersing and stabilising $MoS_2$ in water than SC due to its chemical structure.

We additionally compare the dimensions of the $MoS_2$ flakes exfoliated using SDC and SC using atomic force microscopy (AFM). Samples are prepared by diluting the MoS2 dispersion to 5% v/v and drop-casting onto a $Si/SiO_2$ wafer, before rinsing with DI water to remove the residual surfactant, leaving clean and isolated flakes. The average lateral flake dimensions are very similar for both surfactants, at (90 ± 1) nm for SDC, and (88 ± 1) nm for SC exfoliated $MoS_2$ (Figure 2c). The average flake thickness is smaller in the case of SDC than SC exfoliated $MoS_2$, at (7.8 0.1) nm against (9.6 0.1) nm. Furthermore, the distribution of flake thicknesses is different, with 46% of the SDC exfoliated flakes having thickness <6 layers, compared with 13% of the SC exfoliated flakes (assuming 1 nm measured thickness for a monolayer flake, and 0.7 nm increase for each subsequent layer) [55]. Again, SDC is found to be more effective than SC for $MoS_2$ exfoliation. We thus use SDC assisted $MoS_2$ for composite fabrication.

By combining the measured flake dimensions and dispersed $MoS_2$ concentration, we can verify whether the ex-foliation is limited by the surfactant concentration used. Assuming square flake shapes, it is possible to determine average flake volume, allowing the number of flakes and hence the exfoliated MoS$_2$ surface area in a unit volume of dispersion to be calculated. The calculated surface area of exfoliated MoS2 is 2.6 m$^2$/litre using SC and 5.1 m$^2$/litre using SDC. The concentration of surfactant molecules used in our experiments allows coverage of 3 orders of magnitude greater area, assuming coverage similar to that for SDC on graphene of ~2 nm$^2$/molecule [56], confirming that the exfoliation has not been limited by lack of surfactant molecules.

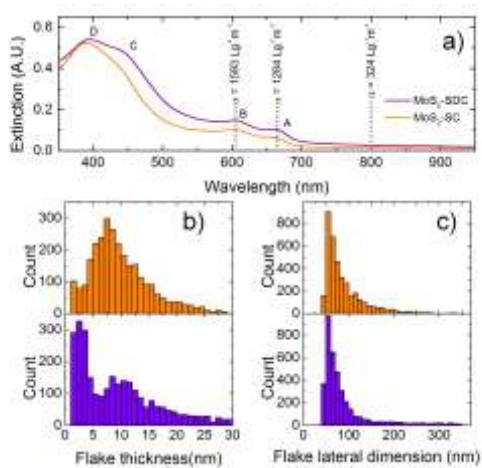

**Figure 2** Characterisation of the exfoliated MoS$_2$. Optical absorption spectrum (a) of the MoS2 dispersions diluted to 10% v/v with background water and surfactant absorption subtracted. The dashed lines indicate the wavelengths used to estimate the concentration. Distribution of flake thicknesses (b) and lateral dimensions (c) measured via AFM for the exfoliated MoS$_2$ (upper panel - SC, lower panel - SDC)

**3 MoS$_2$-polymer composite fabrication and characterisation**

**3.1 MoS$_2$ composite fabrication** For ease of integration into the laser cavity, we use the MoS$_2$ dispersion to fabricate a freestanding polymer composite film, an approach that has previously been successfully used with both 1 and 2d materials [10–12,47]. The composite can then be integrated into a fibre laser by sandwiching between two fibre patch-cords, see Figure 3d. Polyvinyl alcohol (PVA) is widely used for polymer composite SAs due to its solubility in water, as well as the mechanical flexibility and robustness [10,12,47]. We prepare composites of different optical density by mixing 2 mL, 4 mL and 8 mL 0.085 gL$^{-1}$ MoS$_2$ dispersion with an aqueous solution of 15% w/v PVA. For homogeneously dispersed MoS$_2$ flakes in a composite, the optical absorption of the dried film is expected to be equal to that of the dispersion when poured into the petri-dish. This can be estimated from the absorption measurement shown in Figure 2 by considering the relative optical path lengths, allowing us to predict the optical density of our films. The mixture is poured into a petri-dish, and allowed to slowly dry at room temperature in a desiccator, producing a ~25 μm thick freestanding SA. The loading of MoS2 in the dried composites is ~0.06 wt%, ~0.12 wt% and ~0.23 wt% respectively.

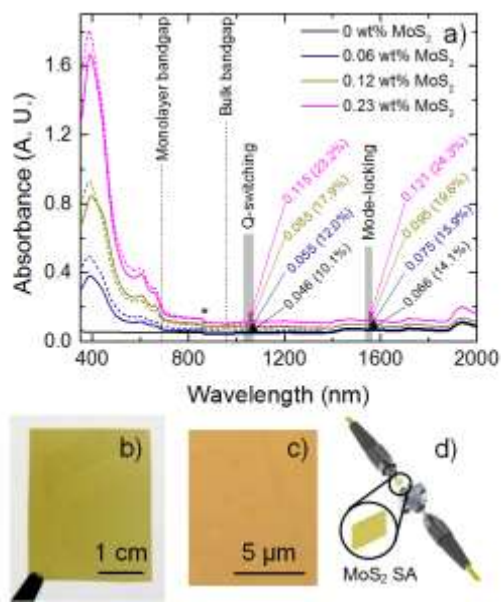

**Figure 3** Characterisation of the MoS$_2$-polymer composite films. Linear optical absorbance of the composite films measured using an integrating sphere to correct for scattering (solid lines) and predicted absorbance (dashed lines) based on the MoS2 content in the films (a). A PVA film of the same thickness is shown for reference. The * indicates a detector change, while the grey boxes show the regions of laser operation. Photograph (b) and optical micrograph (c) of the 0.23 wt% composite film. Integration of the MoS$_2$ SA into the laser cavity (d).

**3.2 MoS$_2$ composite characterisation** The optical absorbance spectra of the composite films are shown in Figure 3a, along with the predicted absorbance according to the MoS$_2$ content in the film. Note that the range of prediction is limited by the high infrared absorption of water, which prevents measurement of the dispersion absorption at wavelengths greater than ~1350 nm. The absorption of the composites is measured using an integrating sphere to rule out any contributions from scattering, particularly in the infrared regions of the spectrum. We observe a good correlation between the predicted and measured absorption, confirming that the absorption of the composite film can be tuned to meet the requirements for a specific laser cavity design. Even at wavelengths below the material bandgap, all of the MoS$_2$ composites show higher absorbance than a reference PVA film of the same thickness. By comparing the relative absorbances of the films, we can see that at 1030-1070 nm, the absorbance from the MoS$_2$ in the composite forms ~16% (0.06 wt%), ~44% (0.12 wt%) and ~56% (0.23 wt%) of the measured absorbance, (~1.9%,

~7.8% and ~13.1% respectively), while at 1535-1565 nm, the absorbance is ~11% (0.06 wt%), ~28% (0.12 wt%) and ~42% (0.23 wt%) of the total (~1.7%, ~5.5% and ~10.2% respectively). For the following sections of this manuscript, we use the composite containing 0.23wt% $MoS_2$, as this would be expected to give the highest modulation depth in the infrared region.

For efficient SA operation, the composite must be free from >1 μm aggregates. This is because such large aggregates could cause non-saturable scattering losses [10]. The stability of the $MoS_2$ dispersion is key to the formation of a uniform composite, as the slow drying process and low viscosity of the starting mixture (~3.8 mPa.s) will otherwise allow aggregation as the $MoS_2$ concentration increases with water evaporation. We use optical microscopy to confirm the absence of such aggregates in our composite; see Figure 3c.

We note that to date, all reports of pulsed lasers using $MoS_2$ SAs have operated at wavelengths longer than 1030 nm, corresponding to photon energies below the $MoS_2$ bandgap [26–35, 37–44]. For large pristine semiconductor crystals, noticeable sub-bandgap absorption is not expected, since the incident photon energy is too low for single photon excitation of electrons into the conduction band. However, as can be seen from the AFM data, our $MoS_2$ flakes have a high edge to surface area ratio (1:11), and we suggest that this sub-bandgap absorption may arise from edge-mediated states in the exfoliated $MoS_2$. Specifically, we propose that the flake edges may lead to a distribution of states within the bandgap, which can be accessed with sub-bandgap excitation energies [28, 36]. Saturation of these energy levels at high incident intensities is followed by Pauli blocking, which enables the material to exhibit absorption saturation [12, 28]. We note that Wang et al have proposed a complementary explanation based on atomic defects causing reduction in the bandgap [37].

**4 Q-switched ytterbium-doped fibre laser** We use the $MoS_2$ composite film as an SA to Q-switch a ytterbium fibre laser. In a Q-switched laser, the SA modulates the Q-factor (i.e. the resonance damping) of the cavity [57]. When the SA is unsaturated, the Q-factor is low, and energy builds up in the cavity. When the SA saturates, the Q-factor of the cavity rapidly increases, causing a high energy pulse output. The energy in the cavity then reduces, and the original state is restored. Q-switched lasers produce relatively long (typically μs-ns) and low repetition rate (kHz) pulses, but with high pulse energy (μJ-mJ) [57], and are therefore often used in applications such as manufacturing, where high energy pulses are required to cure or remove material [6]. The laser setup is shown in Figure 4a. A tunable bandpass filter (1 nm bandwidth) is used to tune the lasing wavelength of the cavity. The laser output is tunable from 1030-1070 nm, with a typical pulse duration of 2.88 μs and a pulse energy of 126 nJ at 1055 nm.

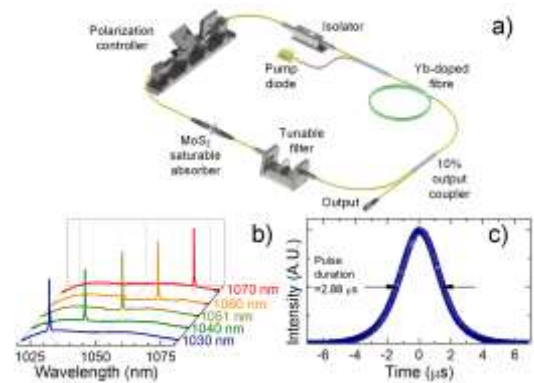

**Figure 4** Tunable fibre laser Q-switched by the $MoS_2$-SA film. Laser setup (a). Output spectra (b) for various wavelengths within the tuning range of 1030-1070 nm. Profile of a single pulse (c).

**5 Mode-locked erbium-doped fibre laser** The $MoS_2$ composite film can also be used to mode-lock an erbium-doped fibre laser. The output of a laser consists of longitudinal modes of the laser cavity, and the intensity undergoes small variations as these move in and out of phase. In a mode-locked laser, the SA is used to couple together some or all of the modes in the laser cavity, such that points in the waveform where the modes are in phase are selectively amplified, while points of antiphase are damped, producing a high repetition rate (MHz-GHz) train of short (ps-fs) pulses, albeit at lower energy than in Q-switched setups [57]. Mode-locked lasers are therefore typically used for applications such as time-resolved spectroscopy [7]. The laser setup is shown in Figure 5a. As before, a wavelength-tunable filter with 12.8 nm bandwidth is used to select the lasing wavelength of the laser cavity. The mode-locked output is continuously tunable from 1535-1565 nm, with typical pulse duration of ~1 ps and pulse energy of 65 pJ.

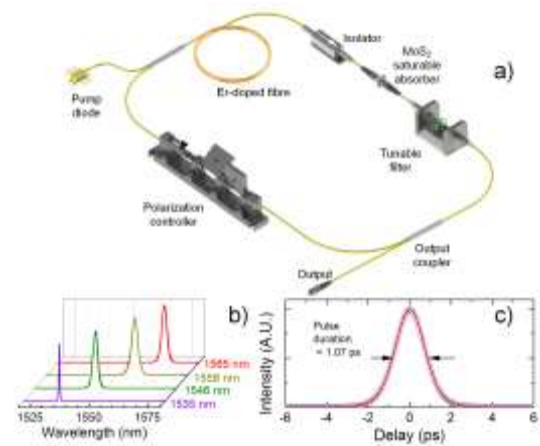

**Figure 5** Tunable fibre laser mode-locked by the $MoS_2$-SA film. Laser setup (a). Output spectra (b) for various wavelengths within the tuning range of 1535-1565 nm. Autocorrelation trace (c) of a single pulse.

**6 Conclusion** We have used surfactant-aided liquid phase exfoliation of $MoS_2$ crystals to produce SAs based on few layer $MoS_2$ flakes. We demonstrated surfactant chemical structures affect the exfoliation and stabilization of $MoS_2$. We can reliably tune the optical density of the $MoS_2$ composite films by varying the $MoS_2$ content. We observe optical absorption from the SA films, even at wavelengths below the $MoS_2$ bandgap. We have used the $MoS_2$ SA to Q-switch an ytterbium-doped fibre laser, tunable from 1030-1070 nm, and additionally to mode-lock an erbium-doped fibre laser, tunable from 1535-1565 nm. We attribute this ability of a single, freestanding $MoS_2$ SA to operate at such a broad range of sub-bandgap wavelengths to edge states in the exfoliated $MoS_2$ nanoflakes. The broad tunability of the operating wavelengths suggests that $MoS_2$ is a very promising material for ultrafast photonic applications.

**Acknowledgements** We thank J. R. Taylor and S. V. Popov for fruitful discussions. EJRK and TH acknowledge support from the Royal Academy of Engineering (RAEng).